\documentclass[9pt,technote]{IEEEtran}

\usepackage{amsmath,amssymb,amsfonts}
\usepackage{algorithmic}
\usepackage{algorithm}
\usepackage{graphicx}
\usepackage{textcomp}
\usepackage{xcolor}
\usepackage{multirow}
\usepackage{booktabs}
\usepackage{cite}
\usepackage{url}
\usepackage{makecell}

\usepackage{hyperref}
\usepackage[caption=false]{subfig} 

\newcommand{\vect}[1]{\boldsymbol{#1}}
\newcommand{\mat}[1]{\mathbf{#1}}
\newcommand{\C}{\mathbb{C}}
\newcommand{\R}{\mathbb{R}}
\newcommand{\expect}{\mathbb{E}}

\title{Unsupervised Online Channel Estimation for High-Mobility OFDM via Implicit Neural Representation}

\author{
	Bohao Shi,~\IEEEmembership{Graduate Student Member, IEEE, }
	Tianfu Qi,~\IEEEmembership{Graduate Student Member, IEEE, }
	Xiaonan Chen,~\IEEEmembership{Member, IEEE, }
	Jun Wang,~\IEEEmembership{Senior Member, IEEE}

	\thanks{

		Bohao Shi, Tianfu Qi, and Jun Wang are with the National Key Laboratory of Wireless Communications, University of Electronic Science and Technology of China, Chengdu, China (e-mail:15032899317@163.com; 202311220634@std.uestc.edu.cn; junwang@uestc.edu.cn).
		
		Xiaonan Chen is with Shenzhen Research Institute of Big Data, The Chinese University of Hong Kong-Shenzhen (e-mail: zs1b200909@gmail.com).
	}
}

\begin{document}
	
	\maketitle
	
	\begin{abstract}
		Accurate channel estimation remains challenging in high-mobility wireless systems because Doppler shifts induce severe inter-carrier interference (ICI) in Orthogonal Frequency Division Multiplexing (OFDM). We propose an unsupervised online channel estimation framework based on Implicit Neural Representation (INR). Unlike discrete-grid estimators, the proposed method decouples channel representation from the OFDM sampling resolution by modeling the time-varying frequency-selective channel as a continuous function of time-frequency coordinates. A Sinusoidal Representation Network (SIREN) with Gaussian Fourier feature mapping captures fine-grained channel variations and high-frequency details without offline pre-training or labeled data. For each received slot, the network parameters are updated by per-slot online fitting that minimizes a physics-aware ICI loss, while a confidence-aware decision-directed loop balances reliable pilots and dynamically harvested pseudo-pilots. Simulations in realistic Vehicle-to-Everything (V2X) environments show that the proposed method achieves near-optimal link-level reliability, significantly outperforming Least Squares (LS) and robust Linear Minimum Mean Square Error (LMMSE) estimators. Compared with supervised deep learning baselines, it also exhibits strong out-of-distribution (OOD) robustness under environmental distribution shifts, establishing an adaptable data-efficient physical-layer paradigm.
	\end{abstract}
	
	
	\section{Introduction}
	\label{sec:intro}

	\IEEEPARstart{O}{rthogonal} Frequency Division Multiplexing (OFDM) remains a core waveform for 5G and emerging 6G systems \cite{He2019Model}, but its reliability depends on accurate channel state information (CSI). In high-mobility Vehicle-to-Everything (V2X) scenarios, rapid channel variation causes Doppler spread and inter-carrier interference (ICI), invalidating the slowly varying diagonal-channel assumption over an OFDM slot \cite{Raviv2024Adaptive, Qiao2025Adaptive}. Classical pilot-aided estimators are difficult to deploy robustly in this regime: Least Squares (LS) amplifies noise and ignores correlation, whereas Linear Minimum Mean Square Error (LMMSE) estimators require second-order channel statistics that are hard to track under non-stationary mobility \cite{Neumann2018Learning, Zhang2020DeepMMSE}.

	These limitations have motivated deep-learning-based estimators. Convolutional Neural Networks (CNNs) can denoise pilot-based estimates, while recurrent and hybrid architectures exploit temporal evolution and mitigate ICI \cite{Soltani2019Deep, Li2023DeepLearningAssisted, Gizzini2023Deep, 10113886}. Such supervised estimators perform well under matched training and deployment channels, but their learned mappings depend on the channel distributions, pilot patterns, mobility levels, and hardware assumptions in the training data. In practical V2X links, these factors may vary jointly with carrier frequency, mobility, scattering profile, pilot density. Distribution mismatch may therefore cause severe OOD performance degradation even when the network architecture itself is sufficiently expressive.

	Reducing dependence on labels and fixed training distributions is therefore essential for high-mobility channel estimation. Model-driven and weakly or unsupervised methods inject physical structure and reduce labeling cost \cite{Bai2019Deep, Qiao2025Adaptive, Tian2024GSURE, Baur2024Leveraging, Baur2024Channel, Zhang2024Unsupervised}. However, per-slot OFDM estimation with scarce pilots requires a stronger form of adaptivity: the estimator should infer the channel directly from the currently received signal without offline CSI labels and without assuming that future slots follow the same distribution as a pre-collected dataset. This requirement motivates an unsupervised online formulation that fits the physical reception model of the current slot rather than learning a static channel-to-channel mapping.

	Implicit Neural Representation (INR) is well suited to this objective. Rather than representing the channel only on a discrete resource grid, INR parameterizes it as a continuous coordinate-to-value function over time and frequency, enabling information sharing across sparse pilots, fine-variation modeling, and evaluation at arbitrary resource elements. This is particularly useful for high-mobility OFDM, where adjacent time-frequency samples remain physically correlated but may exhibit rapid local fluctuations. Although INR has shown promise in wireless channel representation, prediction, location-to-channel mapping, and physics-informed modeling \cite{xiao2022c, chatelier2025model, physics_isac_journal}, this work uses it as a per-slot unsupervised function approximator optimized directly from received OFDM observations.

	We propose SIRIUS (SIREN-based Implicit Representation for Interference-aware Unsupervised Scheme), an unsupervised online estimator for high-mobility OFDM. For each slot, SIRIUS fits a continuous channel function from pilots and reliable decision-directed pseudo-pilots. The function outputs the desired channel response and dominant adjacent ICI coefficients, and is trained by minimizing a physics-aware reconstruction loss derived from the OFDM reception model. Because the parameters are optimized only from the current slot, the estimator neither assumes access to labeled CSI nor requires that future channels follow the statistics of a pre-collected dataset. Online per-slot optimization therefore avoids the dataset-matching requirement of supervised deep estimators under distribution shifts.

	The main contributions of this work are summarized as follows:
	\begin{itemize}
		\item \textbf{Unsupervised online INR for channel estimation:} We formulate high-mobility OFDM channel estimation as per-slot coordinate-based function fitting. The proposed estimator requires neither offline CSI labels nor pretraining on a fixed channel distribution, which improves robustness to OOD channel statistics.
		\item \textbf{ICI-aware continuous channel representation:} To address Doppler-induced ICI, SIRIUS jointly predicts the main channel coefficient and dominant adjacent interference coefficients from time-frequency coordinates. The optimization loss reconstructs the received signal through an explicit OFDM ICI model, giving the learned INR a clear physical interpretation.
		\item \textbf{Confidence-aware decision feedback:} To alleviate pilot scarcity without uncontrolled error propagation, SIRIUS augments true pilots with high-confidence pseudo-pilots selected from hard decisions. A weighted loss assigns stronger trust to true pilots and conservative trust to pseudo-pilots, supporting stable online adaptation.
	\end{itemize}

	\textit{Organization}: Section \ref{sec:system_model} presents the system model, problem formulation, and baselines. Section \ref{sec:method} details SIRIUS. Section \ref{sec:simulation} reports simulation results, and Section \ref{sec:conclusion} concludes the paper.

	\textit{Notations}: Bold uppercase/lowercase letters denote matrices/vectors, e.g., $\mat{X},\mat{Y},\mat{H}$ and $\vect{c},\vect{z},\boldsymbol{\theta}$; $X_{k,n}$ is the $(k,n)$-th entry of $\mat{X}$. Superscripts $(\cdot)^T$ and $(\cdot)^H$ denote transpose and Hermitian transpose, $\hat{(\cdot)}$ denotes an estimate, $\C$ and $\R$ denote complex and real fields, $\Re\{\cdot\}$ and $\Im\{\cdot\}$ denote real and imaginary parts, $\expect[\cdot]$ denotes expectation, and $\mathcal{CN}(0,\sigma^2)$ denotes circularly symmetric complex Gaussian distribution. The operator $\oslash$ denotes element-wise division, $|\cdot|$ denotes modulus, and $[a,b]$ denotes a closed interval for $a\le b$.

	\section{System Model and Problem Formulation}
	\label{sec:system_model}
	
	\subsection{Signal Transmission Model}
	We consider a Single-Input Single-Output (SISO) OFDM system with $K$ subcarriers and $N$ OFDM symbols per slot. Let $\mat{X} \in \C^{K \times N}$ denote the transmitted symbol matrix. The received signal $Y_{k,n}$ at frequency index $k$ and time index $n$ over a time-varying frequency-selective channel is
	\begin{equation}
		Y_{k,n} = H_{k,n}^{(0)} X_{k,n} + \underbrace{\sum_{m \neq 0} H_{k,n}^{(m)} X_{k-m, n}}_{\text{ICI}} + Z_{k,n},
		\label{eq:signal_model_general}
	\end{equation}
	where $H_{k,n}^{(0)}$ is the primary channel response, $H_{k,n}^{(m)}$ is the Inter-Carrier Interference (ICI) coefficient from subcarrier $(k-m)$ to $k$, and $Z_{k,n} \sim \mathcal{CN}(0, \sigma_z^2)$ is additive white Gaussian noise (AWGN). In vector form, $\mat{Y}_n = \mat{H}_n \mat{X}_n + \mat{Z}_n$, where dense matrix $\mat{H}_n$ captures off-diagonal ICI energy.

	The estimation target is the dominant tap set $\{H_{k,n}^{(0)},H_{k,n}^{(-1)},H_{k,n}^{(+1)}\}$ over the full $K\times N$ grid, where $H_{k,n}^{(0)}$ is the desired response and $H_{k,n}^{(\pm1)}$ are adjacent ICI leakage terms used for physical reconstruction. Although higher-order leakage terms may exist in very large Doppler regimes, the adjacent components dominate the local interference pattern in the considered configuration and provide an effective trade-off between model fidelity and online optimization complexity. Under the most severe simulated mobility case ($v=200\,\text{km/h}$ and $f_c=5.9\,\text{GHz}$), the main diagonal together with its two adjacent diagonals contains $99.2\%$ of the effective frequency-domain channel energy, which quantitatively justifies the adopted tri-diagonal ICI representation.

	\subsection{Conventional Channel Estimation}

	Conventional pilot-aided estimators first obtain channel samples at pilot positions and then extend them to the remaining resource elements.
	
	The Least Squares (LS) method estimates pilot-position channels without prior statistics:
	\begin{equation}
		\hat{\mat{H}}_{\mathrm{LS}} = \mat{X}_p^{-1} \mat{Y}_p,
	\end{equation}
	where $\mat{X}_p$ is the diagonal pilot-symbol matrix. LS is simple but suffers from noise enhancement and ICI degradation under high mobility.

	The Linear Minimum Mean Square Error (LMMSE) estimator exploits the channel correlation $\mat{R}_{HH}$ and cross-correlation $\mat{R}_{H\hat{H}_{\mathrm{LS}}}$:
	\begin{equation}
		\hat{\mat{H}}_{\mathrm{LMMSE}} = \mat{R}_{H\hat{H}_{\mathrm{LS}}} (\mat{R}_{HH} + \sigma_z^2 (\mat{X}_p \mat{X}_p^H)^{-1})^{-1} \hat{\mat{H}}_{\mathrm{LS}}.
	\end{equation}
	Although LMMSE is optimal under linear constraints, it requires real-time channel statistics and costly matrix inversions. 

	To reduce this complexity, robust estimators replace the channel-dependent Karhunen-Loeve (K-L) basis in LMMSE with a fixed Discrete Fourier Transform (DFT) basis ($\mat{F}$) \cite{Li1998Robust}:
	\begin{equation}
		\hat{\mat{H}}_{\mathrm{Robust}} = \mat{F} \mat{\Lambda} \mat{F}^H \hat{\mat{H}}_{\mathrm{LS}},
	\end{equation}
	where $\mat{\Lambda}$ is a diagonal noise-suppression matrix. This estimator assumes a worst-case uniform delay profile and Doppler spectrum, but degrades when the instantaneous delay spread or Doppler frequency is substantially different from the predefined design thresholds. Moreover, its fixed transform-domain prior cannot explicitly exploit the instantaneous received data symbols to refine the channel representation within a slot, which limits its adaptability under rapidly changing vehicular channels.

	A representative supervised baseline is the hybrid DNN in \cite{10113886}, which combines a CNN-based channel estimator and a BiGRU-based demodulator. To reduce dimensionality, the CNN refines Basis Expansion Model (BEM) coefficients rather than directly outputting the channel matrix. Let $\hat{\mathbf{h}}_{n}^{\mathrm{DNN}} = \mathcal{F}_{\mathrm{CNN}}(\mathbf{y}_{p,n}, \mathbf{s}_{p,n}, \hat{\mathbf{h}}_{n}^{\mathrm{LS}})$ denote the refined BEM coefficients, where $\mathbf{y}_{p,n}$ and $\mathbf{s}_{p,n}$ are pilot received signals and transmitted symbols, and $\hat{\mathbf{h}}_{n}^{\mathrm{LS}}$ is the coarse LS BEM estimate. The full channel frequency response (CFR) matrix $\hat{\mat{H}}_{n}$ is then analytically reconstructed from $\hat{\mathbf{h}}_{n}^{\mathrm{DNN}}$. This design reduces the output dimension but still relies on offline training data that adequately cover the deployment channel dynamics. To mitigate ICI, a frequency-domain window with radius $B_I$ is formed around subcarrier $k$, and the localized received signal and channel response are processed by a recurrent detector and fused with a windowed LMMSE estimate:
	\begin{equation}
		\mathbf{p}_{k,n}=\mathcal{F}_{\mathrm{Hybrid}}\!\left(\mat{y}_{k-B_I:k+B_I, n}, \hat{\mat{H}}_{k-B_I:k+B_I, n}^{\mathrm{CNN}}, \hat{s}_{k,n}^{\mathrm{LMMSE}}\right),
	\end{equation}
	where $\mathcal{F}_{\mathrm{Hybrid}}$ denotes the BiGRU and fully-connected layers, and $\mathbf{p}_{k,n}$ is the posterior constellation-symbol probability vector. The final decision is $\hat{s}_{k,n}=\arg\max_{s\in\mathcal{S}}\mathbf{p}_{k,n}(s)$. This hybrid design improves tracking under matched conditions but requires large-scale offline labels and may degrade under distribution shifts.

	These limitations motivate a fully online unsupervised estimator.

	\section{Proposed Method}
	\label{sec:method}

	\subsection{Overall Framework}
	
	SIRIUS fits, for each received OFDM slot, a continuous time-frequency function whose outputs are the desired channel tap and dominant adjacent ICI taps. The function is optimized online by reconstructing received OFDM samples under the physical ICI model. Unlike supervised neural receivers, this procedure does not transfer parameters learned from past channel realizations; instead, it uses the neural network as a compact continuous parameterization of the current slot.
	
	High-mobility OFDM channel estimation is under-determined: pilots are sparse, while the channel varies over subcarriers and symbols with non-negligible ICI. Let the unknown coefficients at $(k,n)$ be $\{H_{k,n}^{(0)}, H_{k,n}^{(-1)}, H_{k,n}^{(+1)}\}$. The goal is to recover them over the full $K\times N$ grid from $\mat{Y}$ without offline labeled CSI by minimizing the mismatch between measured and physically reconstructed signals.

	SIRIUS models the channel as a coordinate-to-value mapping $\Phi_{\boldsymbol{\theta}}(\vect{c}_{k,n})$ and updates $\boldsymbol{\theta}$ per slot using only current observations. One shared mapping captures time-frequency channel structure, and grid values are obtained by function evaluation. This formulation is beneficial under sparse pilots because parameters are shared globally across coordinates rather than estimated independently at each resource element.

	The mapping explicitly couples the continuous coordinate input with the complex channel quantities to be reconstructed. It is defined as
	\begin{equation}
	\begin{split}
	\Phi_{\boldsymbol{\theta}}(\vect{c}_{k,n})
	&= \Bigg[
	\Re\{\hat{H}_{k,n}^{(0)}\},\ 
	\Im\{\hat{H}_{k,n}^{(0)}\},\ \cdots \\ &
	\Re\{\hat{H}_{k,n}^{(-1)}\},\ 
	\Im\{\hat{H}_{k,n}^{(-1)}\},\ 
	\Re\{\hat{H}_{k,n}^{(+1)}\},\ 
	\Im\{\hat{H}_{k,n}^{(+1)}\}
	\Bigg]^T,
	\end{split}
	\label{eq:sirius_output_mapping}
	\end{equation}
	where the six real outputs are the real and imaginary parts of the main tap ($\Re\{\hat{H}_{k,n}^{(0)}\}, \Im\{\hat{H}_{k,n}^{(0)}\}$) and two adjacent ICI taps ($\Re\{\hat{H}_{k,n}^{(\pm1)}\}, \Im\{\hat{H}_{k,n}^{(\pm1)}\}$).

\begin{figure}[!t]
		\centering
		\includegraphics[width=\linewidth]{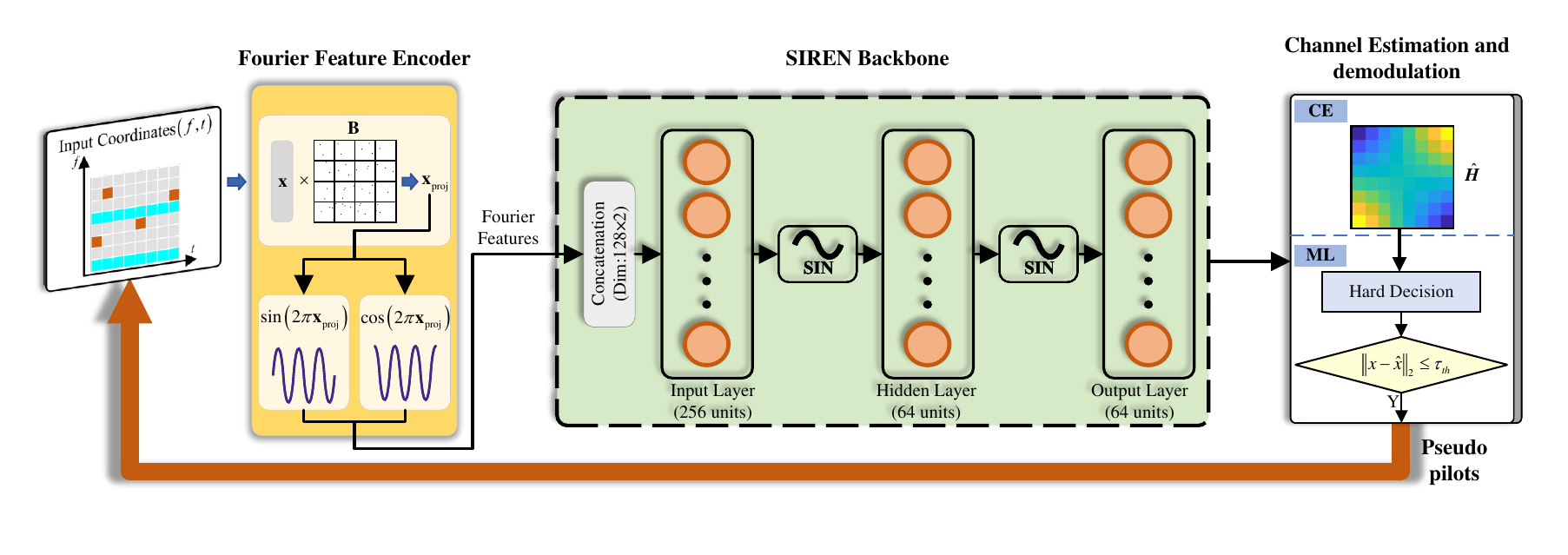} 
		\caption{Schematic of the proposed SIRIUS framework, illustrating the coordinate-based INR mapping and the iterative decision-directed feedback loop.}
		\label{fig:framework_final}
	\end{figure}

	As shown in Fig.~\ref{fig:framework_final}, SIRIUS converts sparse pilots and reliable pseudo-pilots into a slot-wise continuous channel function through iterative warm-start, decision, and retraining.
	
	Given $\mat{Y}$, pilot positions, and pilot symbols, LS initialization on pilot subcarriers provides a coarse estimate $\hat{\mat{H}}_{\mathrm{init}}$. The equalization and decision module produces tentative data symbols and confidence scores; high-confidence decisions are merged with true pilots as pseudo-pilots for the next optimization round. The INR module maps time-frequency coordinates to updated channel taps, which are fed back to equalization until the final channel estimate is obtained for symbol detection. This closed-loop design converts reliable decisions into additional supervision while keeping low-confidence symbols excluded from the training set.

	\subsection{Network Architecture: Fourier-Enhanced SIREN}
	
	The architecture implements \eqref{eq:sirius_output_mapping} with input coordinate $\vect{c}_{k,n} = [f_k, t_n]^T \in [-1, 1]^2$.

	\subsubsection{Gaussian Fourier Feature Mapping}

	Fourier features reduce the spectral bias of coordinate MLPs, since raw coordinates may over-smooth rapidly varying high-mobility channels. This issue is especially important in the considered OFDM setting because Doppler-induced time selectivity and delay-induced frequency selectivity jointly create localized high-frequency structures across the resource grid.
	
	According to Neural Tangent Kernel (NTK) theory \cite{jacot2018neural}, standard coordinate-based MLPs converge faster to low-frequency components than to high-frequency details. In high-mobility V2X channels, large Doppler shifts and multipath delay spreads cause rapid temporal and frequency selectivity; thus, raw coordinates $\vect{c}_{k,n}=[f_k, t_n]^T$ limit fast-fading tracking. Following \cite{tancik2020fourier}, we project the input into a higher-dimensional periodic space before the SIREN backbone using a Gaussian Random Fourier Feature (RFF) mapping $\gamma: \mathbb{R}^2 \rightarrow \mathbb{R}^{2M}$:
	 \begin{equation}
	 	\gamma(\vect{c}_{k,n})=\left[\cos(2\pi \mat{B}\vect{c}_{k,n})^T, \sin(2\pi \mat{B}\vect{c}_{k,n})^T\right]^T,
	\end{equation}
	where $\mat{B} \in \mathbb{R}^{M \times 2}$ has independent entries drawn from $\mathcal{N}(0, \sigma^2)$. This projection transforms the NTK into a shift-invariant interpolation kernel and shifts spectral bias toward a high-frequency band determined by $\mat{B}$. The scale factor $\sigma$ controls the resolvable time-frequency bandwidth; empirically, $\sigma=0.5$ provides a suitable spectral spread for Doppler- and delay-induced channel variations, enabling stable online convergence and accurate reconstruction.

	\subsubsection{SIREN Backbone}
	The encoded features pass through sine-activated layers with layer-specific frequency multipliers for gradient propagation and convergence across frequencies. Sine activations are adopted because periodic basis functions can represent oscillatory channel variations more naturally than piecewise-linear activations in coordinate-based implicit models. The $l$-th hidden layer is
	\begin{equation}
		\vect{z}^{(l)} = \sin\left(\omega^{(l)} (\mat{W}^{(l)} \vect{z}^{(l-1)} + \vect{b}^{(l)})\right),
	\end{equation}
	where $\omega^{(l)}$ is $\omega_{\mathrm{first}} = 30.0$ for $l=1$ and $\omega_{\mathrm{hidden}} = 30.0$ for $l > 1$.

	The final linear layer outputs the six real components of $\{\hat{H}_{k,n}^{(0)},\hat{H}_{k,n}^{(-1)},\hat{H}_{k,n}^{(+1)}\}$ in Eq.~\eqref{eq:sirius_output_mapping}.

	\subsection{Iterative Confidence-Aware Optimization}

	SIRIUS performs per-slot optimization through a confidence-aware decision-feedback loop. Let $\Omega_{\mathrm{p}}$ and $\Omega_{\mathrm{d}}$ denote the true-pilot and selected pseudo-pilot positions, respectively. The active set is $\mathcal{D}=\Omega_{\mathrm{p}}\cup\Omega_{\mathrm{d}}$, with weights $w_{\mathrm{p}}>w_{\mathrm{d}}$ assigned to pilots and pseudo-pilots.
	
	Since full-grid CSI labels are unavailable, the network is trained by received-sample reconstruction. In each outer iteration of Algorithm \ref{alg:online_estimation}, SIRIUS updates the INR parameters, equalizes the received symbols with the current $\hat{H}^{(0)}$, and admits a hard decision as a pseudo-pilot only when its Euclidean distance to the nearest constellation point is below $\tau_{\mathrm{th}}=0.5$. The self-supervised objective is
	\begin{equation}
	\begin{aligned}
		\mathcal{L}(\boldsymbol{\theta}) = &\frac{1}{N_{\mathcal{D}}} \sum_{(k,n) \in \mathcal{D}} w_{k,n} \left| Y_{k,n} - \hat{Y}_{k,n} \right|^2 \\
		&+ \frac{\lambda}{KN} \sum_{k=0}^{K-1} \sum_{n=0}^{N-1} \left( \left|\hat{H}^{(-1)}_{k,n}\right|^2 + \left|\hat{H}^{(+1)}_{k,n}\right|^2 \right),
	\end{aligned}
	\end{equation}
	where $N_{\mathcal{D}}$ is the number of active samples and $\hat{Y}_{k,n}=\hat{H}_{k,n}^{(0)}X_{k,n}+\hat{H}_{k,n}^{(-1)}X_{k-1,n}+\hat{H}_{k,n}^{(+1)}X_{k+1,n}$. Exact pilots anchor the reconstruction, while high-confidence pseudo-pilots densify supervision. The regularization term suppresses localized noise fitting by penalizing adjacent ICI taps over the full $K\times N$ grid.

	This feedback mechanism requires a minimally reliable initial estimate. At extremely low SNR, e.g., below $0\,\text{dB}$, LS-based initialization may produce many erroneous decisions, causing error propagation through $\Omega_{\mathrm{d}}$. Confidence thresholding and reduced pseudo-pilot weights mitigate but cannot eliminate this risk; harsher conditions may require conservative admission, soft-decision weighting, or channel-code assistance.

	\begin{algorithm}[!t]
		\caption{SIRIUS: Iterative Channel Estimation via Physics-Aware INR}
		\label{alg:online_estimation}
		\begin{algorithmic}[1]
			\REQUIRE {Received signal matrix $\mat{Y}$, pilot set $\Omega_{\mathrm{p}}$, pilot symbols $X_{k,n}$ for $(k,n)\in\Omega_{\mathrm{p}}$}
			\ENSURE {Full channel-tap estimates $\{\hat{H}^{(0)},\hat{H}^{(-1)},\hat{H}^{(+1)}\}$}
			\STATE Set stabilization constant $\epsilon=10^{-8}$ and confidence threshold $\tau_{\mathrm{th}}=0.5$
			
			\STATE \textbf{Warm Start:} LS estimate on pilots + Interpolation $\to \hat{\mat{H}}_{\mathrm{init}}$
			\STATE Initial Equalization: $\hat{X}_{k,n}=\text{HardDec}(Y_{k,n}/(\hat{H}_{\mathrm{init},k,n}+\epsilon))$, $\forall(k,n)$
			\STATE {Initialize $\Omega_{\mathrm{d}}\leftarrow\emptyset$, $\mathcal{D}\leftarrow\Omega_{\mathrm{p}}$, and $w_{k,n}=w_{\mathrm{p}}$ for $(k,n)\in\Omega_{\mathrm{p}}$}
			
			\FOR{Outer Iteration $i = 1$ \TO $I_{\max}=2$}
				\STATE \textit{// Phase 1: Network Training}
				\STATE Set gradient steps $S_i=150$ for $i=1$ and $S_i=50$ for $i=2$
				\FOR{Gradient Step $s = 1$ \TO $S_i$}
					\STATE Compute data loss over $\mathcal{D}$ using weights $w$
					\STATE Compute sparsity regularization over the full $K \times N$ grid
					\STATE $\mathcal{L}(\boldsymbol{\theta}) \leftarrow \text{Data Loss} + \lambda \cdot \text{Regularization}$
					\STATE Update network parameters: $\boldsymbol{\theta} \leftarrow \boldsymbol{\theta} - \eta \nabla_{\boldsymbol{\theta}} \mathcal{L}$
				\ENDFOR
				
				\STATE \textit{// Phase 2: Inference \& Equalization (if $i < I_{\max}$)}
				\IF{$i < I_{\max}$}
					\STATE Predict primary tap $\hat{H}_{k,n}^{(0)} \leftarrow \Phi_{\boldsymbol{\theta}}(\text{All } \vect{c}_{k,n})$
					\FOR{each subcarrier $k$ and symbol $n$}
					\STATE Equalize signal: $\tilde{X}_{k,n} = Y_{k,n} / (\hat{H}_{k,n}^{(0)} + \epsilon)$
					\ENDFOR
					
					\STATE \textit{// Phase 3: Pseudo-Pilot Harvesting}
					\STATE Hard decision: $\hat{X}_{k,n} = \text{HardDec}(\tilde{X}_{k,n})$
					\STATE Calculate confidence distance: $d_{k,n} = \left| \tilde{X}_{k,n} - \hat{X}_{k,n} \right|$
					\STATE Update $\mathcal{D}$: Append points where $d_{k,n} < \tau_{\mathrm{th}}$ with $w=w_{\mathrm{d}}$
				\ENDIF
			\ENDFOR
			\STATE Predict Final Triplets $\{\hat{H}_{k,n}^{(0)}, \hat{H}_{k,n}^{(-1)}, \hat{H}_{k,n}^{(+1)}\} \leftarrow \Phi_{\boldsymbol{\theta}}(\text{All } \vect{c}_{k,n})$
			\RETURN $\hat{\mat{H}} = \{\hat{H}^{(0)}, \hat{H}^{(-1)}, \hat{H}^{(+1)}\}$
		\end{algorithmic}
	\end{algorithm}	
	
	\section{Simulation Results and Analysis}
	\label{sec:simulation}

	SIRIUS is evaluated in a 5G NR V2X SISO-OFDM environment using the TDL-C (Urban Canyon) profile \cite{3gpp38901} with RMS delay spread $93\,\text{ns}$. At $5.9\,\text{GHz}$ and $100(200)\,\text{km/h}$, the maximum Doppler shift is $546.3(1092.6)\,\text{Hz}$, yielding a strongly ICI-limited condition. This setup stresses both the temporal tracking capability and the robustness of the channel estimator to frequency leakage. System configurations and SIRIUS hyperparameters are summarized in Tables \ref{tab:sim_params} and \ref{tab:network_params}, respectively.

	\begin{table}[htbp]
		\centering
		\caption{System Simulation Parameters}
		\label{tab:sim_params}
		\renewcommand{\arraystretch}{1.2}
		\begin{tabular}{lc}
			\toprule
			\textbf{Parameter} & \textbf{Value} \\
			\midrule
			Carrier Frequency / SCS & $5.9\,\text{GHz}$ / $30\,\text{kHz}$ \\
			FFT Size / Active Subcarriers & $512$ / $288$ \\
			Slot Structure & 14 OFDM Symbols \\
			Pilot Pattern & Comb-type \\
			Pilot Interval & $8$ \\
			Max Doppler Shift & \makecell{$546.3\,\text{Hz}$ ($\approx 100\,\text{km/h}$) \\ 
$1092.6\,\text{Hz}$ ($\approx 200\,\text{km/h}$)} \\
			Modulation & QPSK \\
			Channel Model & TDL-C (Urban Canyon) \\
			\bottomrule
		\end{tabular}
	\end{table}

	\begin{table}[!t]
		\centering
		\caption{Neural Network and Algorithm Configuration}
		\label{tab:network_params}
		\resizebox{\columnwidth}{!}{
			\begin{tabular}{ccc}
				\toprule
				\textbf{Module} & \textbf{Parameter} & \textbf{Value} \\
				\midrule
				\multirow{2}{*}{Input} & Coord Dims ($f, t$) & 2 \\
				& Normalization & $[-1, 1]$ \\
				\midrule
				\multirow{3}{*}{Fourier Mapping} & Mapping Size $M$ & 128 \\
				& Output Dim & 256 ($\sin + \cos$) \\
				& Scale Factor $\sigma$ & $0.5$ \\
				\midrule
				\multirow{4}{*}{Backbone} & Hidden Dim & 64 \\
				& Layers & 4 \\
				& First Freq. $\omega_{\mathrm{first}}$ & 30.0 \\
				& Hidden Freq. $\omega_{\mathrm{hidden}}$ & 30.0 \\
				\midrule
				Output & Dimension & 6 (3x Complex) \\
				\midrule
				\multirow{5}{*}{Optimization} & Pilot Weight $w_{\mathrm{p}}$ & 2.0 \\
				& Pseudo-Pilot Weight $w_{\mathrm{d}}$ & 0.5 \\
				& Decision Threshold $\tau_{\mathrm{th}}$ & 0.5 \\
				& Sparsity Penalty $\lambda$ & 1.0 \\
				& Learning Rate $\eta$ (Adam) & $5\times10^{-4}$ \\
				\bottomrule
			\end{tabular}
		}
	\end{table}

	All experiments were conducted on an Intel(R) Core(TM) i9-14900HX CPU at $2.20\,\text{GHz}$ with $32.0\,\text{GB}$ RAM and an NVIDIA GeForce RTX 4060 Laptop GPU with $8\,\text{GB}$ memory. Table \ref{tab:runtime} reports the measured per-slot runtime, offline training cost, and trainable model size under the same OFDM configuration and hardware platform. The closed-form LS, ideal LMMSE, and robust LMMSE estimators achieve the lowest latency, while the Hybrid DNN incurs moderate inference cost and requires offline training. In contrast, SIRIUS eliminates offline pre-training and uses only $0.029$ M trainable parameters, but its per-slot optimization results in a measured runtime of $0.7\,\text{s}$ on the tested GPU. This latency is still longer than the physical duration of a 14-symbol 5G NR slot with $30\,\text{kHz}$ subcarrier spacing; hence, strict online deployment would require substantial FPGA/ASIC acceleration. Incorporating meta-learning-based initialization is a promising way to retain online adaptability while reducing the number of fine-tuning steps required for each slot, thereby lowering the per-slot training delay.

	\begin{table}[!t]
		\centering
		\caption{Measured Runtime, Training Cost, and Model Size of Compared Algorithms}
		\label{tab:runtime}
		\renewcommand{\arraystretch}{1.2}
		\resizebox{\columnwidth}{!}{
			\begin{tabular}{lccc}
				\toprule
				\textbf{Algorithm} & \makecell{\textbf{Online Runtime} \\ \textbf{per Slot (s)}} & \makecell{\textbf{Offline Training} \\ \textbf{Time}} & \makecell{\textbf{Trainable} \\ \textbf{Parameters}} \\
				\midrule
				LS & 0.0065 & N/A & N/A \\
				Ideal LMMSE & 0.0066 & N/A & N/A \\
				Robust LMMSE & 0.0069 & N/A & N/A \\
				Hybrid DNN \cite{10113886} & 0.2785 & 6 min 15.43 s & 0.977 M \\
				SIRIUS & 0.7 & N/A & 0.029 M \\
				\bottomrule
			\end{tabular}
		}
	\end{table}

	\subsection{Evaluation of the Proposed Architecture}

	\subsubsection{Impact of Fourier Feature Mapping} 
\begin{figure}[htbp]
		\centering
		\subfloat[]{%
			\includegraphics[width=0.7\linewidth]{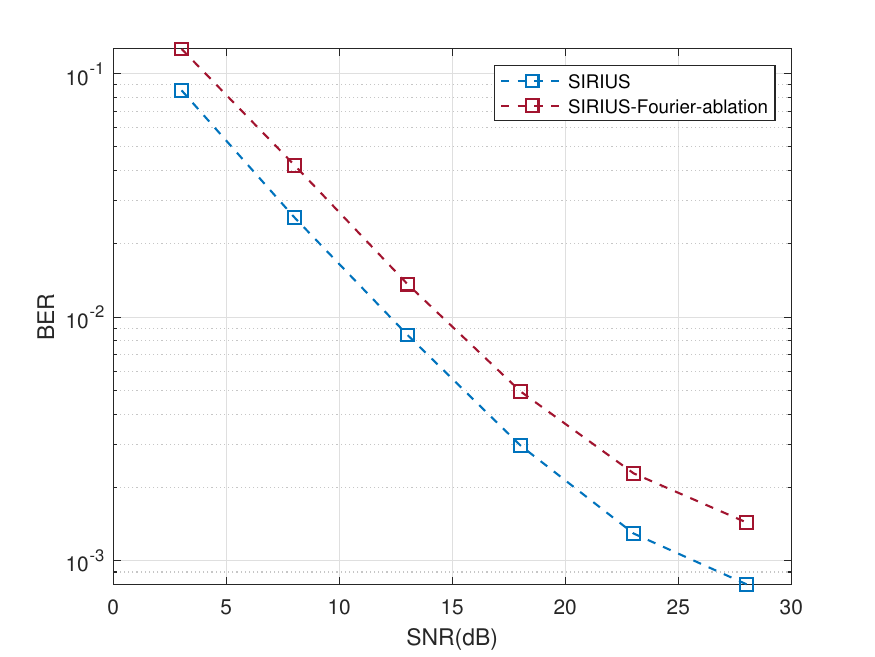}%
			\label{fig:sub1}%
		}
		\vspace{-12pt}  
		\subfloat[]{%
			\includegraphics[width=0.7\linewidth]{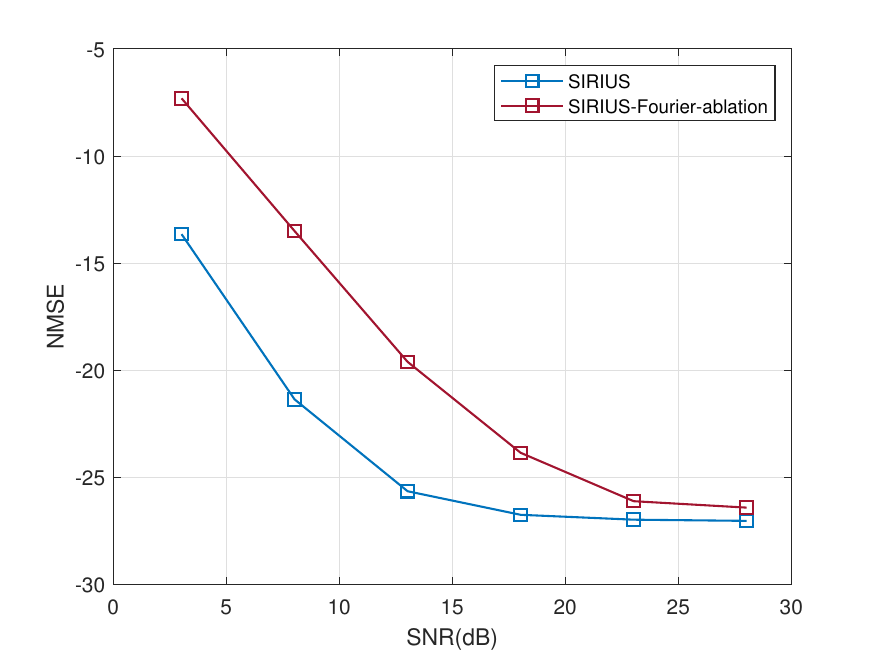}%
			\label{fig:sub2}%
		}
		\caption{BER and NMSE performance for evaluating the impact of the Fourier feature mapping.}
		\label{fig:ablation_ff}
	\end{figure}

Structural ablations validate the SIRIUS architecture. Fig. \ref{fig:ablation_ff} shows that introducing the 256-dimensional Fourier feature projection improves the NMSE mainly in the low-to-medium SNR regime, while the high-SNR NMSE gain becomes relatively limited. This saturation is caused by residual modeling error under sparse pilot supervision and decision-directed pseudo-pilot uncertainty; once noise is sufficiently suppressed, these structural factors dominate the estimation error and lead to an NMSE floor. In contrast, the BER is improved over the whole SNR range because the Fourier mapping provides a richer representation of rapid time-frequency channel variations and yields more reliable equalized symbols even when the NMSE improvement becomes marginal.

	\subsubsection{Convergence of Iterative Optimization} 
\begin{figure}[!t]
		\centering
		\subfloat[]{%
			\includegraphics[width=0.7\linewidth]{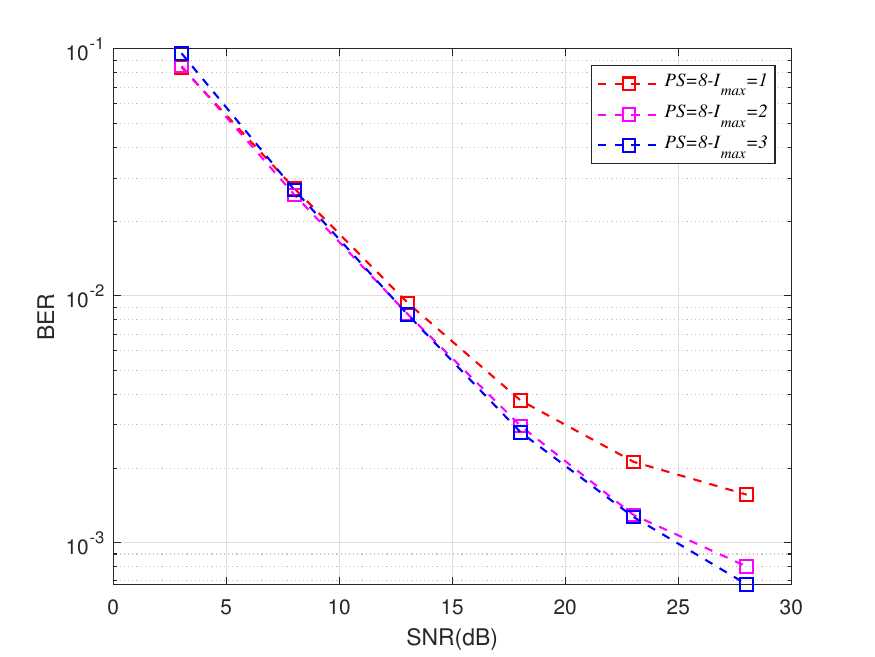}
			\label{fig:iter_BER}
		}
		\vspace{-12pt}  
		\subfloat[]{%
			\includegraphics[width=0.7\linewidth]{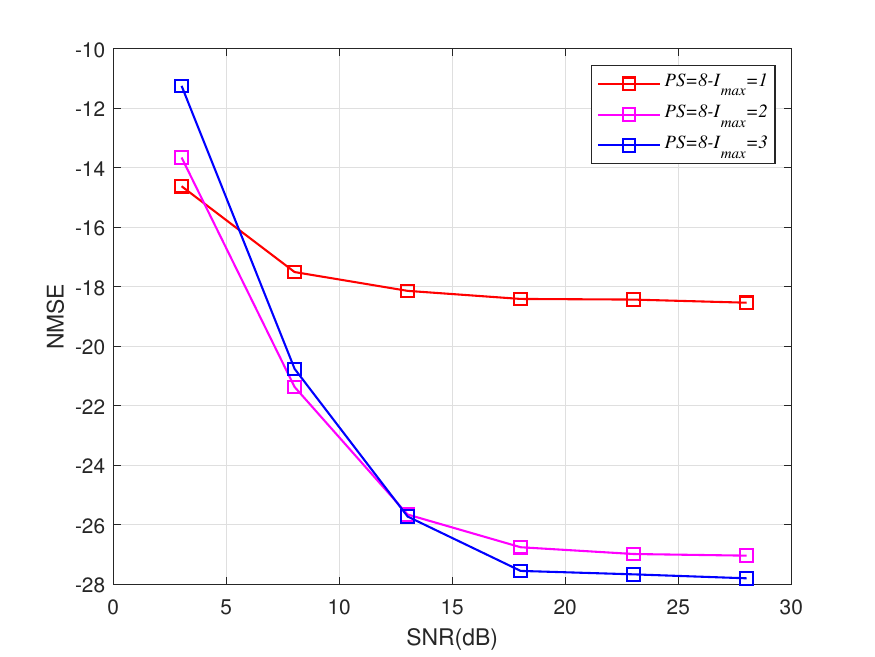}
			\label{fig:iter_NMSE}
		}
		
		\caption{BER and NMSE performance of SIRIUS under different numbers of outer-loop iterations.}
		\label{fig:iteration_impact}
	\end{figure}

Fig. \ref{fig:iteration_impact} relates estimation accuracy to the number of outer-loop iterations ($I_{\max}$). One decision-directed feedback pass ($I_{\max}=2$) markedly reduces NMSE from the initial state, while further iterations yield only marginal gains, indicating rapid saturation. Even with pilot spacing 16, $I_{\max}=2$ provides the best accuracy-complexity trade-off, showing that one pseudo-pilot harvesting stage is sufficient for dense supervision. This observation is important for online deployment because it limits the retraining overhead while preserving the performance benefit of decision feedback.

	\subsection{Comparison with Previous Approaches}
    SIRIUS is benchmarked against LS, ideal LMMSE, robust LMMSE with fixed worst-case statistics \cite{Li1998Robust}, and the supervised hybrid DNN \cite{10113886}. For fair channel-tracking comparison, Zero-Forcing (ZF) equalization is used for symbol detection across all estimators. 
    
    The hybrid DNN uses the same pilot arrangement as SIRIUS and is trained offline at $25\text{ dB}$ SNR. Its CNN-based estimator fuses four-dimensional pilot features with coarse LS estimates via an MLP, and its BiGRU equalizer captures temporal dependencies within a sliding window of radius $B_I=5$ (a $11$-symbol observation interval). The sequential features are concatenated with a local LMMSE prior for final symbol classification. This baseline is representative of supervised high-mobility receivers because it combines channel refinement and data detection, but it also illustrates the dependence of learned receivers on the training distribution.

\begin{figure}[!t]
        \centering
        \subfloat[]{%
            \includegraphics[width=0.7\linewidth]{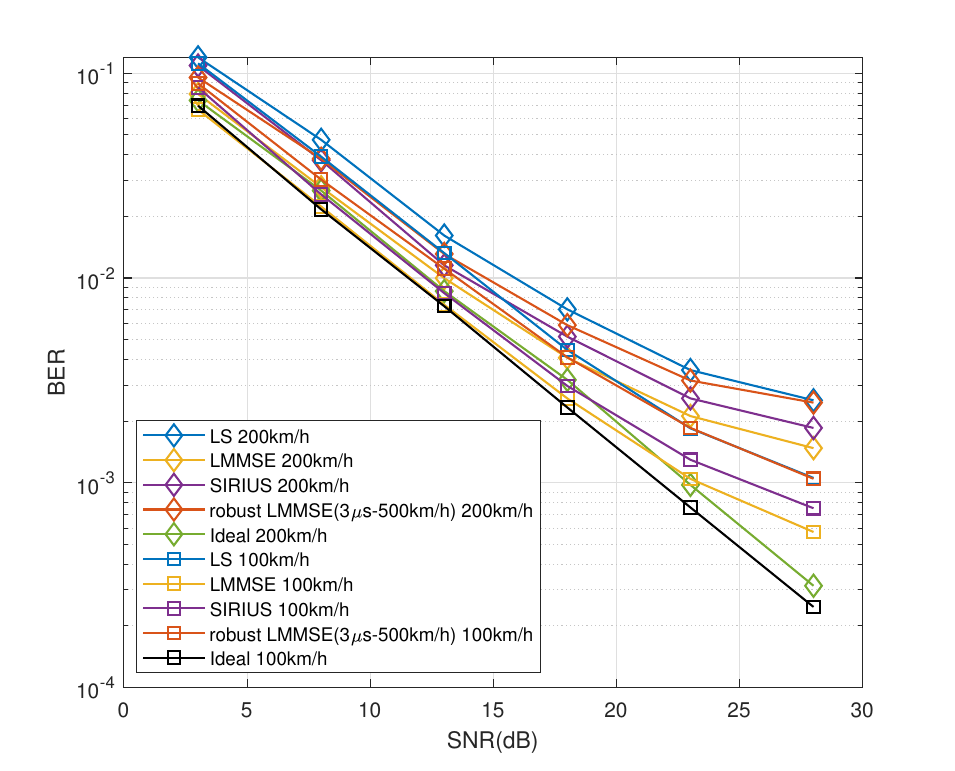}
            \label{fig:sota_ber}
        }
        \vspace{-12pt}  
        \subfloat[]{%
            \includegraphics[width=0.7\linewidth]{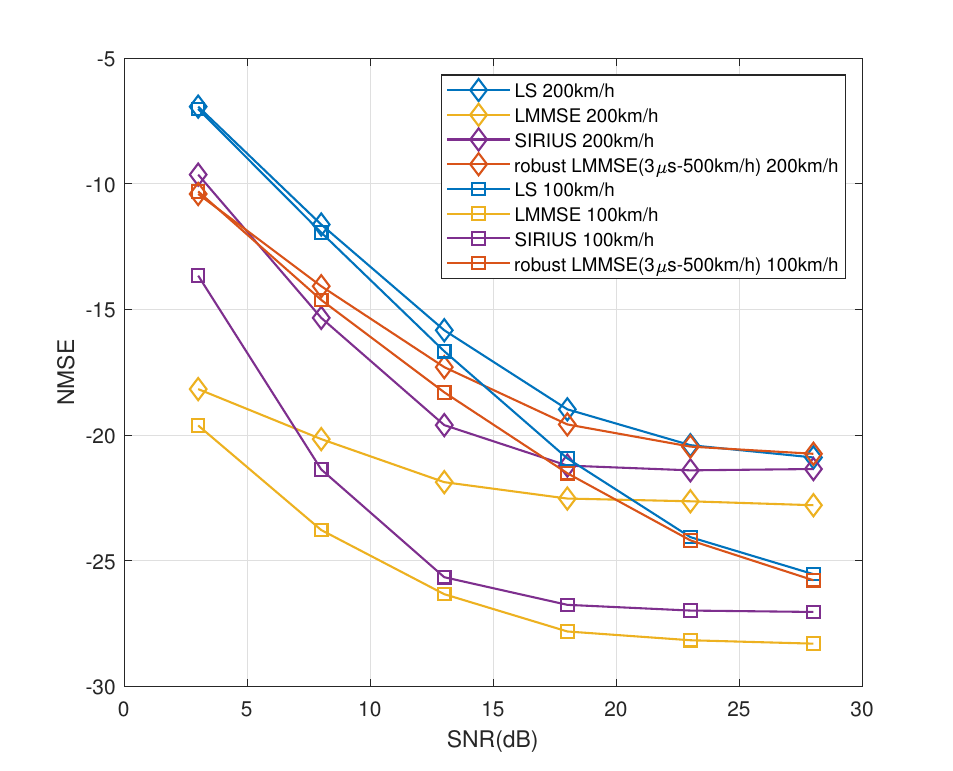}
            \label{fig:sota_nmse}
        }
        \caption{BER and NMSE performance comparison among LS, ideal LMMSE, robust LMMSE, and the proposed SIRIUS using Zero-Forcing (ZF) equalization.}
        \label{fig:LS-LMMSE-RLMMSE-SIRIUS}
    \end{figure}

    Fig. \ref{fig:sota_ber} shows uncoded BER, with the theoretical lower bound obtained under perfect CSI and ideal ICI cancellation. In the low-to-medium SNR regime ($\text{SNR} < 23\text{ dB}$), robust LMMSE outperforms LS; however, at $\text{SNR} \ge 23\text{ dB}$, its fixed worst-case assumptions (maximum delay spread $3\ \mu\text{s}$ and velocity $500\text{ km/h}$) cause convergence to the LS error floor. SIRIUS maintains a consistent advantage and ranks second only to ideal LMMSE. In the $100\text{ km/h}$ case, at BER $10^{-3}$, SIRIUS incurs only $1.8\text{ dB}$ degradation relative to ideal LMMSE and gains $3\text{ dB}$ over robust LMMSE and LS. Under the more severe $200\text{ km/h}$ mobility, SIRIUS requires about $4\text{ dB}$ higher SNR than ideal LMMSE to reach the same BER level, while still achieving approximately $5\text{ dB}$ gain over robust LMMSE and LS.

	Fig. \ref{fig:sota_nmse} further reports NMSE under both mobility levels. Ideal LMMSE sets the benchmark with perfect prior statistics, whereas LS suffers from noise amplification and ICI neglect. By avoiding the statistical inflexibility of robust LMMSE and explicitly learning the dominant adjacent ICI taps, SIRIUS achieves high estimation accuracy. In the $100\text{ km/h}$ case, at NMSE $-26\text{ dB}$, SIRIUS incurs only a $2\text{ dB}$ penalty relative to ideal LMMSE and provides a $14\text{ dB}$ gain over robust LMMSE and LS. In the $200\text{ km/h}$ case, at NMSE $-21\text{ dB}$, SIRIUS requires about $7\text{ dB}$ higher SNR than ideal LMMSE, while achieving approximately $10\text{ dB}$ gain over robust LMMSE and LS. These results confirm that the proposed online INR fitting can approach the ideal statistical estimator without requiring real-time channel covariance information.

\begin{figure}[!t]
        \centering
        \subfloat[]{%
            \includegraphics[width=0.7\linewidth]{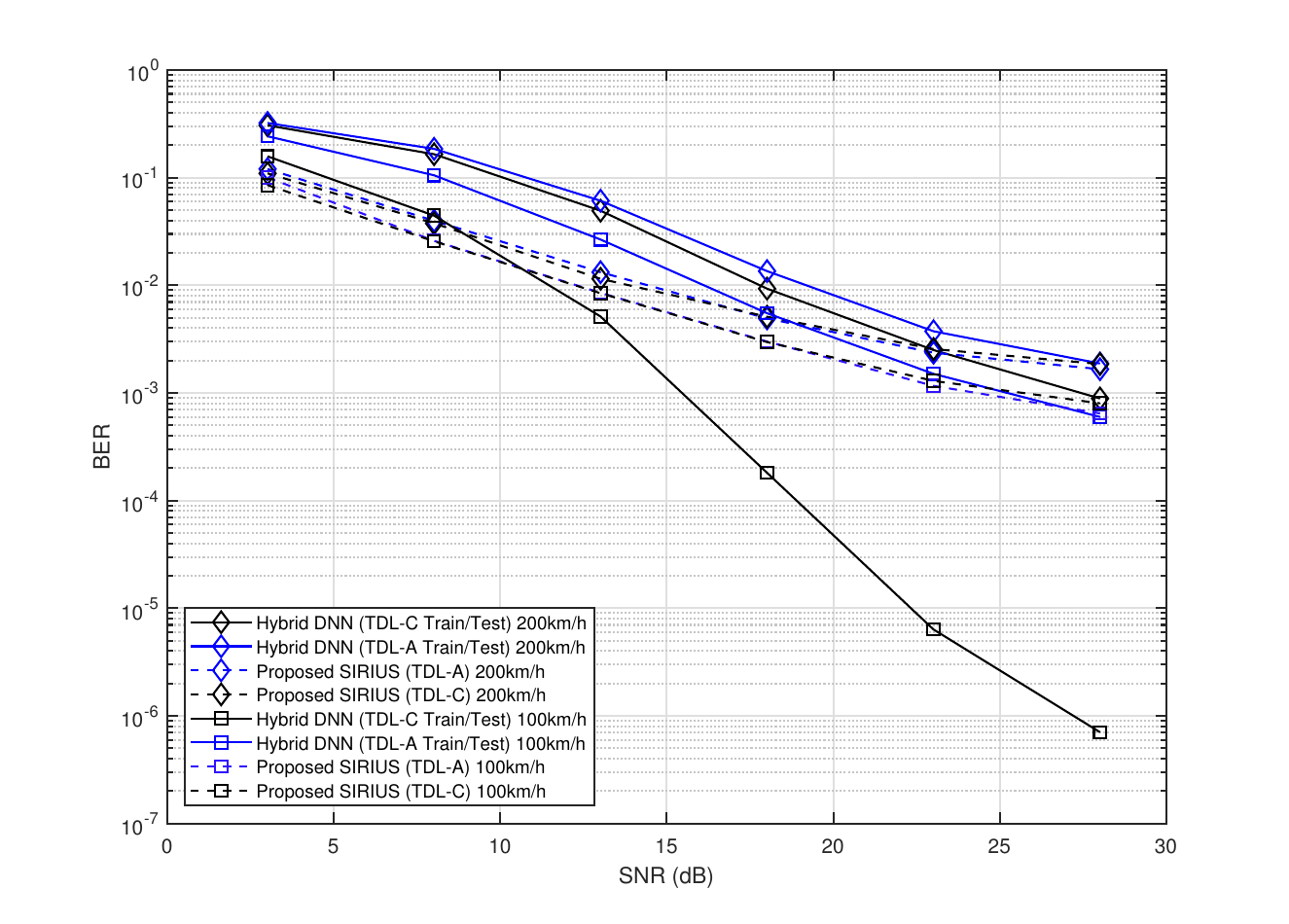}
            \label{fig:dnn_ber}
        }
        \vspace{-12pt}  
        \subfloat[]{%
            \includegraphics[width=0.7\linewidth]{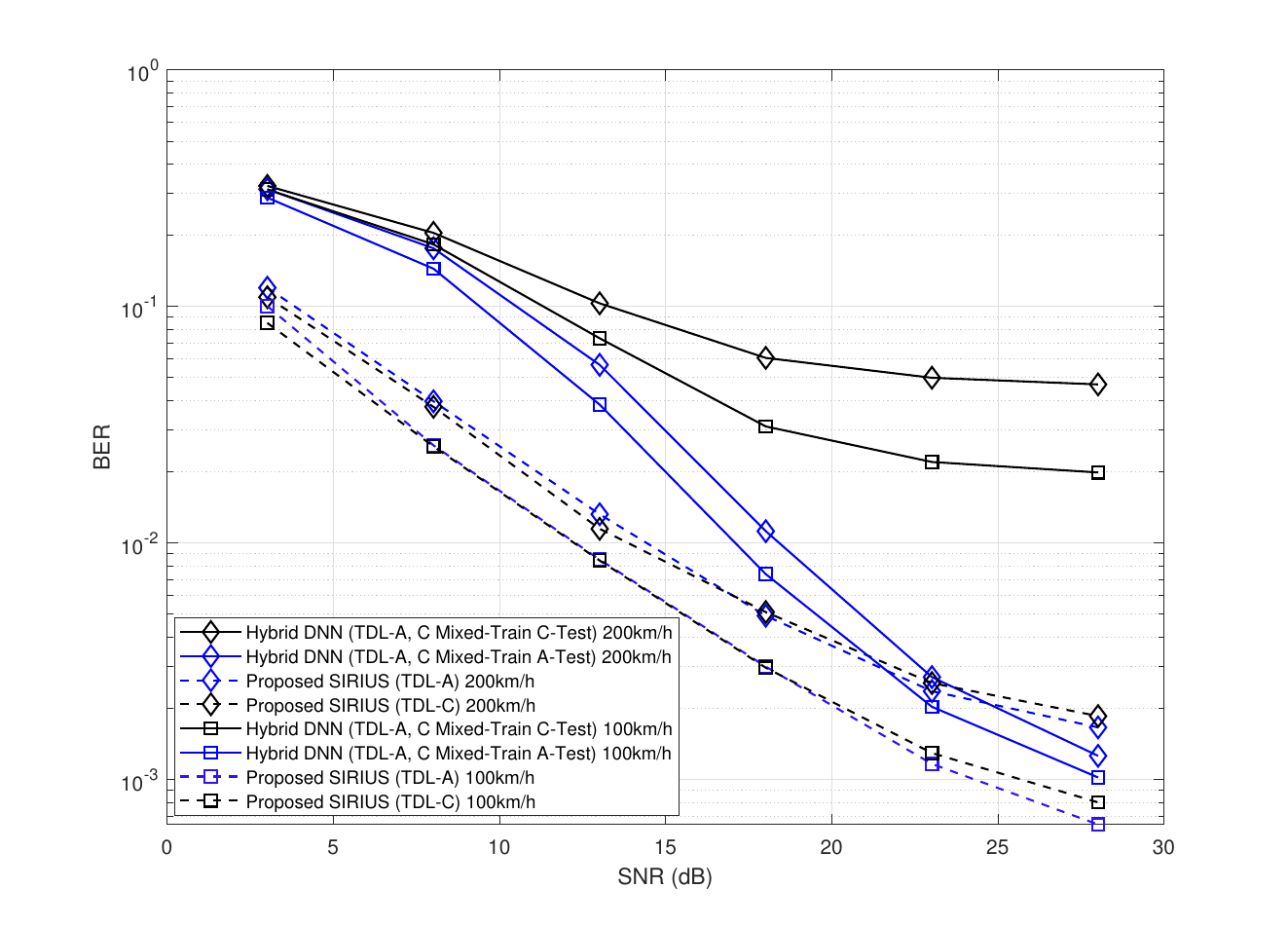}
            \label{fig:dnn_gen}
        }
        \caption{Performance comparison between the proposed SIRIUS and the supervised hybrid DNN architecture \cite{10113886} in terms of (a) BER under scenario-matched conditions (TDL-A and TDL-C) and (b) generalization capabilities under mismatched training and testing conditions.}
        \label{fig:dnn_comparison}
    \end{figure}

    SIRIUS is also compared with the supervised hybrid DNN \cite{10113886} in Fig. \ref{fig:dnn_comparison}, covering uncoded BER under scenario-matched conditions (Fig. \ref{fig:dnn_ber}) and generalization under distribution shifts (Fig. \ref{fig:dnn_gen}). This comparison separates the ability to learn a specialized mapping for a fixed scenario from the ability to adapt to unseen channel statistics, which is essential for V2X deployment.
    
    Fig. \ref{fig:dnn_ber} shows consistent SIRIUS detection performance across TDL-A and TDL-C, whereas the hybrid DNN is scenario-dependent. When trained and tested on TDL-C, the DNN learns a customized nonlinear mapping and outperforms SIRIUS; when trained and tested on TDL-A, it is inferior to SIRIUS. Thus, although the hybrid DNN can excel with sufficient data and tuning for a specific setting, changing the dataset does not ensure the same superiority. This behavior is consistent with the fact that the supervised model encodes statistical regularities of the training channel, while SIRIUS re-optimizes its representation from the observations of each received slot. 
    
    This limitation becomes pronounced when training and testing distributions differ, as in dynamic V2X networks. Fig. \ref{fig:dnn_gen} shows that the offline-trained hybrid DNN degrades when trained on a mixed dataset comprising TDL-A and TDL-C profiles and evaluated under TDL-A and TDL-C test conditions, due to limited OOD generalization. SIRIUS exhibits stronger robustness and sustains reliable links under varying vehicular conditions without offline retraining. These results indicate that per-slot physics-aware optimization provides better adaptation than a fixed supervised mapping when the propagation environment is not known in advance.

    \section{Conclusion}
    \label{sec:conclusion}
    This paper proposed SIRIUS, an unsupervised online channel estimator for high-mobility OFDM systems. By combining a Fourier-enhanced SIREN with a physics-aware ICI reconstruction loss and confidence-guided decision feedback, SIRIUS learns the time-frequency channel directly from each received slot without offline CSI labels or scenario-specific pre-training. Simulation results under realistic V2X channels demonstrate that SIRIUS approaches the ideal LMMSE benchmark, outperforms LS and robust LMMSE, and exhibits stronger robustness than supervised hybrid DNNs under distribution shifts. The current implementation still incurs non-negligible online optimization latency, and its first-order ICI truncation is best suited to the considered Doppler range. Future work will investigate meta-learning-based initialization, hardware acceleration, and adaptive ICI-window selection to reduce latency and extend applicability to more extreme mobility regimes.
	
	\bibliographystyle{IEEEtran}
	\bibliography{references_zh}
	
\end{document}